\documentstyle[procl,epsf]{article}

\input{psfig}

\def\Journal#1#2#3#4{{#1} {\bf #2}, #3 (#4)}

\def\NPB{{\em Nucl. Phys.} B}

\def\PLB{{\em Phys. Lett.}  B}

\def\PRD{{\em Phys. Rev.} D}

\def\ZPC{{\em Z. Phys.} C}

\def\be{\begin{equation}}

\def\ee{\end{equation}}

\def\bea{\begin{eqnarray}}

\def\eea{\end{eqnarray}}

\newcommand{\gev}{\mbox{$\;$GeV}}

\newcommand{\PP}{I\!\!P}

\begin{document}

\begin{flushright}
University of Edinburgh 96/15\\
University of Liverpool LTH-378\\
hep-lat/9607373
\end{flushright}

\title{SOFT POMERON PHYSICS ON THE LATTICE}

\author{D.S. HENTY, D.G. RICHARDS, J.I. SKULLERUD }

\address{Department of Physics and Astronomy, University of Edinburgh,
Mayfield Road,\\
Edinburgh EH9 3JZ, Scotland}

\author{C. PARRINELLO}

\address{Department of Mathematical Sciences, University of Liverpool,\\
Liverpool L69 3BX, England}

\author{(UKQCD Collaboration)}

\maketitle\abstracts{We use 
 lattice QCD to investigate 
some topics in strong interaction phenomenology which 
are usually related to soft pomeron exchange. 
These include hadronic cross-sections at high energies and diffractive 
scattering at HERA. 
Some numerical results are presented in the framework of the
Landshoff-Nachtmann pomeron model, and strategies for 
further studies are discussed.}

\section{Introduction}

It has been known for a long time
 that the behaviour of the total hadronic 
cross section at high energies could be described in 
the framework of Regge theory.~\cite{Collins} 
In particular, the rise of the cross section 
was attributed to the Regge trajectory with the quantum numbers of the vacuum:
 the Pomeron.~\cite{land_DIS}
Renewed 
interest in Pomeron 
models has been recently triggered by the analysis of diffractive 
scattering at HERA. In particular, the discovery of events with large rapidity 
gaps has been interpreted as evidence for Pomeron 
exchange.~\cite{HERA1}$^{\!-\,}$\cite{HERA2}

Recovering the 
phenomenology of Pomeron exchange from first principles QCD 
is a major theoretical problem. 
The lattice formulation of QCD, together with the Monte Carlo method, provides 
an invaluable tool for nonperturbative, first principles 
calculations. The
 aim of the present note is to describe some attempts to use such a tool 
in order
 to test models of diffraction and pomeron exchange from 
the 
point of view of QCD. 
In the following we first recall some basic facts about lattice 
QCD, then we describe an application of the method to  
test the Landshoff-Nachtmann (LN) model,~\cite{LN} and finally
 we outline a general approach to the 
lattice investigation of diffractive physics. 

\section{Lattice QCD}

The transition from the continuum formulation of QCD to the lattice one 
can be summarised as follows:~\cite{Creutz} 
first the theory is continued to imaginary times ($t \rightarrow i t$) and
 4-dimensional spacetime is mapped onto the sites of a (usually 
hypercubic) lattice. Then 
 a gauge-invariant discretisation of the QCD Lagrangian is obtained 
after defining fermion fields $\psi (n), \bar{\psi} (n)$ 
at the lattice sites and $SU(3)-$valued 
parallel transport operators $U (n,n^{'})$, connecting 
nearest-neighbour sites. The relation between the link variables $U$ and 
the continuum gauge potential $A$ is given by $U = exp(i a g A)$, where $a$ 
is the lattice spacing and $g$ is the bare coupling constant.
In this process two fundamental length scales are introduced: the lattice 
spacing $a$, which acts as an ultraviolet cutoff, and the volume of the box 
$V=L^4$, which regulates the infrared behaviour. Thus the above procedure  
produces a gauge-invariant, nonperturbative regularisation of the (Euclidean) 
theory. 
Finally, path integrals yielding Green functions of quark, gluon and hadronic 
fields can be numerically computed by means of the Monte Carlo method. 

In the continuum limit, the dependence of the bare coupling constant on 
the value of $a$ is dictated 
by the renormalisation group equations. Asymptotic freedom implies 
$g_0 \rightarrow 0$ when $a \rightarrow 0$. 
In practice, the continuum limit $a \rightarrow 0$ can be studied 
numerically by 
letting $g_0$ approach zero. 
In such a limit physical quantities 
will not depend on the value of the cutoff, while 
ultraviolet-divergent Green functions will, and  
suitable renormalisation conditions can be imposed.
 Because of computer power limitations, most lattice QCD studies, 
including the present one,  
are still performed in the {\it quenched approximation}, which amounts to 
neglecting quark loops. 
Unlike in the continuum formulation of the theory,
 gauge-invariant, physical quantities can be directly computed on the lattice 
without gauge fixing. However, for our purposes one also needs to compute  
gauge-dependent Green functions. In this case a consistent gauge-fixing 
procedure has to be implemented. For the studies discussed here 
the lattice version of the  
Landau gauge was used.

\section{The Landshoff-Nachtmann Pomeron on the Lattice} 

In the LN model 
 Pomeron exchange between quarks behaves like a
$C=+1$ photon-exchange
diagram, with amplitude
\begin{equation}
i \beta^2_0 (\bar{u} \gamma_\mu u)(\bar{u} \gamma^\mu u).
\end{equation}
$\beta_0$ represents the strength of the Pomeron coupling to quarks,
and is related to the (non-perturbative) gluon propagator by
\begin{equation}
\beta_0^2 = \frac{1}{36 \pi^2} \int d^2 p \left[ g^2 D(p)\right]^2,
\label{eq:beta0_constraint}
\end{equation}
where $g$ is the gluon-quark coupling.
$\beta_0$ can be
determined from, for example, the total $pp$ cross
section.~\cite{landshoff:84}
As a consequence, the model yields simple formulae for $pp$ scattering,
exclusive $\rho$ production in deep inelastic scattering and the $J/ \Psi - $
nucleon total cross section, which all contain integrals in momentum space of
$g^2 D(p)$.~\cite{halzen:93,ducati:93}

In order to extract predictions from the model, one needs an expression
for $g^2 D(p)$.
Obviously, convergence of (\ref{eq:beta0_constraint}) requires
that the infra-red pole of the gluon propagator be removed by some
nonperturbative mechanism.
By inserting in the LN model the gluon propagator
 as computed on the lattice, we aimed to perform 
a strong consistency test 
from the point of view of QCD.~\cite{noi} 
We performed calculations on two sets of hypercubic lattices, corresponding to
different physical volumes and lattice spacings. 
In addition, we used data
for the gluon propagator as evaluated by Marenzoni 
{\it et al.}.~\cite{stella:94} 
Comparing results from different data sets is important in 
order to study 
possible lattice artifacts in our
calculations.
We extracted the scalar gluon self-energy $D_{\rm lat} (p)$ from our data
 and then 
fitted it to various analytical expressions. Then the curves corresponding 
to the best fits
were inserted in the LN formulae.
In this process, we assumed that in the continuum limit
the propagator
is multiplicatively renormalisable, as it is in perturbation theory.
Also, we neglected the running of the QCD coupling, i.e.
we made the approximation $g(p) = g$.
As the scale for the momenta in $D_{\rm lat} (p)$ was set
from independent measurements, we only had one
free parameter  to fix in the expression $g^2 D_{\rm lat} (p)$.
This is a multiplicative factor that
corresponds to the product of a gluon wavefunction renormalisation constant
times a numerical value for $g^2$.
We call this parameter $g_{\rm eff}^2$.
It can be determined by using (\ref{eq:beta0_constraint}) as a
nonperturbative 
renormalisation condition, i.e. by imposing that $\beta_0$ attains
its phenomenological value of 2.0 ${\gev}^{-1}$:
\begin{equation}
\beta_{0}^2 = \frac{1}{36 \pi^2} \int d^2 p \left[ g_{\rm eff}^2 \ D_{\rm lat} (p)
\right]^2 = 4 \ {\gev}^{-2}
\label{eq:beta0_fix}
\end{equation}
It is the combination $g_{\rm eff}^2 D_{\rm lat} (p)$ that we insert in the
formulae of the LN model. In using such formulae, 
we adopt the analysis procedure of Halzen and 
collaborators~\cite{halzen:93,ducati:93}.
 
\subsection{Proton-proton elastic scattering}
 
The calculation of $\sigma^0_{\rm tot}$ and $\frac{d \sigma^0}{dt}$, i.e. 
the energy-independent part of the 
total and elastic differential cross section for proton-proton
scattering, provides a benchmark for the LN model of the Pomeron.
The model has already been explored using nonperturbative
propagators obtained from approximate solutions of 
Schwinger-Dyson equations.~\cite{halzen:93}
Single Pomeron exchange is expected to dominate the
differential cross-section up to $-t \simeq 0.5 \, {\gev}^2$.  To fully
describe the energy dependence, the intercept of the Pomeron
trajectory is taken to be somewhat larger than 1. The measured
total and elastic 
differential cross sections are usually parametrised as
follows:
\begin{equation}
\sigma_{\rm tot}  =  \left(\frac{s}{m_p^2}\right)^{0.08}
\sigma^0_{\rm tot}, \qquad 
\frac{d \sigma}{dt}  =  \left(\frac{s}{m_p^2}\right)^{0.168}
\frac{d \sigma^0}{dt}. 
\label{eq:dsig_energy}
\end{equation}
For small $t$, the elastic differential cross section behaves like
$e^{B t}$, and the model is characterised by two parameters,
$\sigma^0_{\rm tot}$ and $B$.
 
We computed
$\sigma^0_{\rm tot}$ and $B$ on each lattice using the lattice
gluon propagator and the effective coupling $g_{\rm eff}$.
We obtained values of $\sigma^0_{\rm tot}$ ranging from  $18.12 \ {\rm mb}$ to 
$19.85 \ {\rm mb}$ 
and
$B$ from $12.6 {\gev}^{-2}$ to $13.6 {\gev}^{-2}$, thus
in very good agreement with 
each other, suggesting 
that both quantities are subject to only small discretisation and 
finite volume effects.
 They are also encouragingly close to the phenomenological
values of $\sigma^0_{\rm tot} \simeq 22.7\,{\rm mb}$ and $B \sim
11~\gev^{-2}$.
\begin{figure}
\begin{center}
\leavevmode
\epsfysize=250pt
\epsfbox[20 30 620 600]{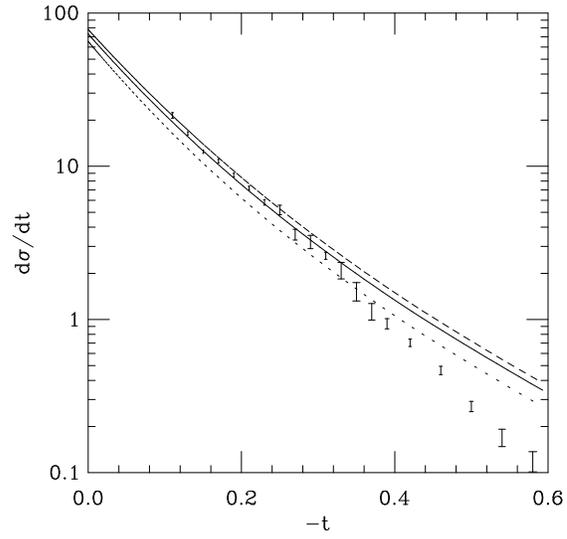}
\end{center}
\caption{Data~\protect\cite{isr:84}
 for the $pp$ elastic cross section at $\protect\sqrt{s} =
53~\gev$ together with the lattice
predictions, corrected for the energy dependence, on the $16^4$ lattices at
$\beta = 6.2$ (solid) and $\beta = 6.0$ (dots), and on the $24^3
\times 32$ lattice at $\beta=6.0$ (dashes).}
\label{fig:isr_data}
\end{figure}
In Fig.~\ref{fig:isr_data} we show ISR data~\protect\cite{isr:84}
 for the differential
elastic cross section at $\sqrt{s} = 53~\gev$ together with the
lattice predictions, with the energy correction of
Eq.~\ref{eq:dsig_energy}. 
 
\subsection{$J/\psi$-Nucleon Scattering}
This process provides a further important test of the LN model.
\begin{figure}
\begin{center}
\leavevmode
\epsfysize=250pt
\epsfbox[20 30 620 600]{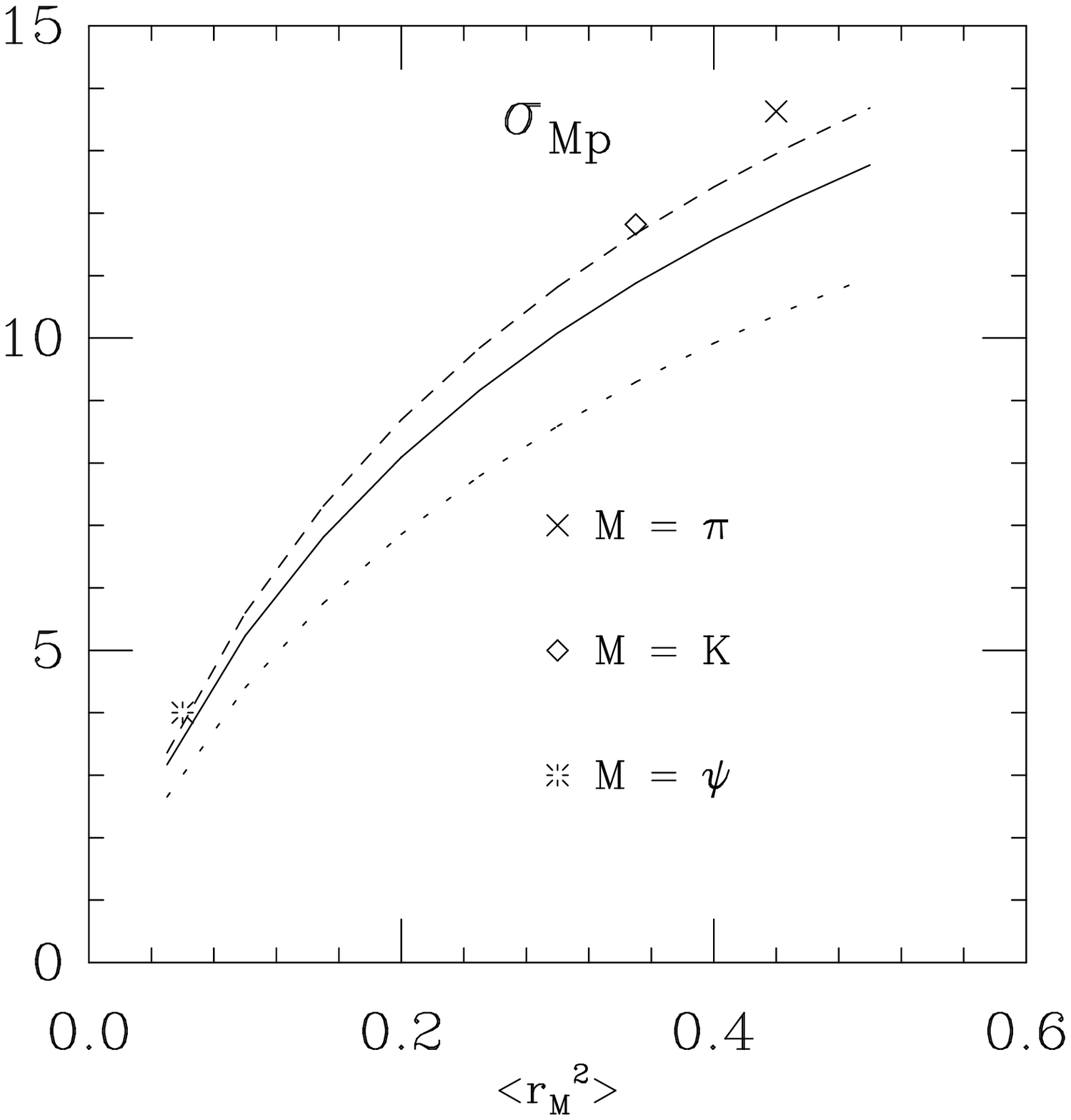}
\end{center}
\caption{The meson-nucleon total cross section as a function of the
radius used in the meson form factor, on
the $16^4$ lattices at $\beta=6.2$ (solid) and $\beta=6.0$ (dots), and
on the $24^3 \times 32$ lattice at $\beta=6.0$ (dashes).  Also shown
are the radii and total cross sections corresponding to the $\pi$, $K$
and $J/\psi$.}
\label{fig:jpsi}
\end{figure}
Two phenomenological features emerge from our calculations.  Firstly, the
quark-counting rule is closely satisfied in the case of the hadrons
composed of light quarks, as we get for the ratio $\sigma_{pp}/\sigma_{\pi p}$ 
values in the range $1.5 - 1.8$.
Secondly, the Pomeron couples more weakly to mesons composed of
heavier quarks.  This can be seen in
Fig.~\ref{fig:jpsi},
 where we show the meson-nucleon cross section as a function
of the pole radius, together with a Regge
fit to the energy-independent part of the $\pi^{-} p$ and $K^{-} p$
cross sections.~\cite{landshoff:92} However,  we observe 
now sizeable differences
between the results on the different lattices, reinforcing the need to
repeat the calculation for a wider range of lattice parameters.
  
\section{Diffractive Physics: the Factorisation Hypothesis}

In this section we summarise some attempts to develop 
a general approach for investigating diffractive
scattering 
using lattice QCD.
Our approach starts from assuming 
{\it factorisation}.  
To clarify this concept, consider diffractive scattering at HERA in the 
one-photon approximation:
\begin{equation}
\gamma^*(q)+p(p)\to\tilde p(\tilde p)+X(p_X).
\label{eq:scat}
\end{equation}

Following Arens.{\it et al.},~\cite{landlast} we say that 
if the above process factorises, one can describe it in two steps:
first the original proton emits a Pomeron,  
$p(p) \rightarrow p(\tilde p)+ \PP(\Delta)$,
then the Pomeron collides with
the $\gamma^*$ to produce the final hadronic state $X$,
$\gamma^*(q)+ \PP (\Delta) \rightarrow X(p_X)$.
In other words, the amplitude for (\ref{eq:scat}) can be written in a
factorizable form:
\begin{equation}
A(\gamma^*p \rightarrow X p)=A(\gamma^*\PP \rightarrow X)\otimes(\PP-
{\rm propagator})\otimes(pp\PP-{\rm vertex}).
\end{equation}
Notice that the above assumption does not imply that we treat the Pomeron like
 a 
particle. Also, we emphasise that although the physics we discuss here 
is usually interpreted in terms of Pomeron exchange, we want to analyse it 
from a more general point of view, where we only assume factorisation 
and the emission of a colour singlet excitation. In this sense, our use of 
the word Pomeron in the present discussion is simply for illustrative 
purposes. 
On the lattice we want to study the second and third factors in the above 
formula, i.e. Pomeron propagation and the 
effective coupling of the Pomeron to a proton. One motivation for such 
a study is the investigation of the helicity structure of the Pomeron, as it
 has 
recently been argued that a nontrivial spin structure in the 
Pomeron-proton vertex may be related to quantities measurable at 
HERA.~\cite{landlast,golo} 
 Given some {\it ansatz} for a composite QCD operator 
$O_{\PP} (x)$ which creates the Pomeron from the QCD vacuum, the effective 
Pomeron-proton coupling  takes the form of a QCD 3-point vertex function, which we can study in momentum space as a function of the momentum of the proton, 
$p$, and the momentum of the Pomeron, $\Delta$.
Both because of technical limitations and because of theoretical 
prejudice, we will 
assume that $O_{\PP} $ only contains gluonic fields. 
Usually composite QCD operators are classified according to $J^{PC}$, i.e. 
spin, parity and charge conjugation. Since the Pomeron is a Regge trajectory, 
all the information we have {\it a priori} is that $O_{\PP}$ should have 
a $J^{++}$ structure.
As we do not want to assign a value to $J$, it makes sense to 
allow $O_{\PP}$ to have arbitrary Lorentz structure, i.e. to be the sum
of a scalar, a vector, and tensors of arbitrary rank. 

Numerical work is in progress, aiming to test different {\it ans\"atze} 
for $O_{\PP}$. 
We are currently focusing on two-gluon operators, and we have 
obtained encouraging results as far as the numerical feasibility of the 
project is concerned.

\section*{Acknowledgments}
This work was supported by the United
Kingdom Particle Physics and Astronomy Research Council (PPARC) under
grant GR/J21347.  CP and DGR acknowledge the support of PPARC through
Advanced Fellowships held at the University of Liverpool and at the
University of Edinburgh respectively.

\end{document}